\documentclass[aps,prl,twocolumn,groupedaddress]{revtex4}

\usepackage{graphics} 
\usepackage{amsmath}
\usepackage{verbatim}
\newcommand{\um}{\ensuremath{\mu\textrm m}}

\begin{document}


\title{Realization of a feedback controlled flashing ratchet}



\author{Benjamin J. Lopez}
\email[]{blopez1@uoregon.edu}
\affiliation{Department of Physics and Materials Science Institute, University of Oregon, Eugene, OR 97403-1274}

\author{Nathan J. Kuwada}
\affiliation{Department of Physics and Materials Science Institute, University of Oregon, Eugene, OR 97403-1274}

\author{Erin M. Craig}
\affiliation{Department of Physics and Materials Science Institute, University of Oregon, Eugene, OR 97403-1274}

\author{Brian R. Long}
\affiliation{Department of Physics and Materials Science Institute, University of Oregon, Eugene, OR 97403-1274}

\author{Heiner Linke}
\email[]{linke@uoregon.edu}
\affiliation{Department of Physics and Materials Science Institute, University of Oregon, Eugene, OR 97403-1274}


\date{\today}

\begin{abstract}

A flashing ratchet  transports diffusive particles using a time-dependent, asymmetric potential. Particle speed is predicted to increase when a feedback algorithm based on particle positions is used. We have experimentally realized such a feedback ratchet using an optical line trap, and observed that use of feedback increases velocity by up to an order of magnitude. We compare two different feedback algorithms for small particle numbers, and find good agreement with simulations. We also find that existing algorithms can be improved to be more tolerant to feedback delay times.  

\end{abstract}

\pacs{}

\maketitle

\begin{figure}[!b]
\includegraphics{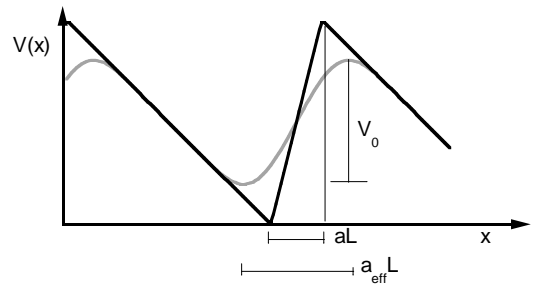}
\caption{\label{ratchetdiagram}Black: a piece-wise linear ratchet potential $V_{lin}(x)$ with period $L$ and asymmetry $a = 0.2$. Gray: The effective potential $V_{\textrm{eff}}(x)$ (Eq.~\eqref{effective}) experienced by a sphere of radius $L/2$ trapped in $V_{\textrm{lin}}(x)$ has a reduced asymmetry $a_{\textrm{eff}} = 0.34$, and a reduced potential height $V_{0}$.} 
\end{figure}

Brownian motors \cite{Astumian:2002} induce directed motion of diffusive particles and are used as models for biological and artificial nanomachines. A flashing ratchet \cite{Astumian:1994fk,Prost:1994uq} is a specific type of Brownian motor that uses an asymmetric ratchet potential $V(x)$, with spatial period $L$ (Fig. \ref{ratchetdiagram}) that can be turned on (control parameter $\alpha  = 1$) or off ($\alpha  = 0$). If the potential is turned on and off randomly or periodically on the time scale of particle diffusion over the distance $aL$ (Fig. \ref{ratchetdiagram}) the mean particle velocity can reach up to the order of $L/2$ per flashing cycle. 

Recent theory work predicts that the performance of a flashing ratchet can be dramatically improved by feedback algorithms that control $\alpha (t)$ based on particle distributions \cite{Cao:2004lr}. Experimental implementation of feedback control in ratchets is also predicted to exhibit synchronization effects \cite{Craig:2008ys, Feito:2007vn} and current reversals \cite{Feito:2008rr}, and could be used to model paradoxical games \cite{Parrondo:2004} or gating mechanisms that allow processivity in linear, dimeric molecular motors \cite{Bier:2004}. One specific feedback algorithm maximizes the instantaneous velocity (MIV) of overdamped particles by setting the control parameter $\alpha$ such that the net force on the particles is always zero or positive \cite{Cao:2004lr}: 
	
	\begin{equation}
	\label{MIV}
	\alpha  = \Theta \left( {\sum\limits_i^N {F\left( {x_i} \right)} } \right)       \  \textrm{(MIV)}.
	\end{equation}

Here  $\Theta$ is the Heaviside function, $F(x_{i})$ is the force on each particle if the ratchet potential were turned on, and $x_{i}$ is the particle position. For small $N$ this scheme results in a time-averaged center of mass velocity, $<\!\!v\!\!>$, an order of magnitude larger than in an optimally operated, periodically flashing ratchet \cite{Cao:2004lr}.

Recently, another feedback scheme was proposed that predicts a moderate improvement on the performance of the MIV scheme for small $N\! > \!1$ \cite{Craig:2008fr}. This scheme maximizes the net displacement (MND) of the particles that is expected when the ratchet potential is turned on:
	
	\begin{equation}
	\label{MND}
	\alpha  = \Theta \left( {-\sum\limits_i^N {\left( {x_i  - x_0 } \right)} } \right) \ \textrm{(MND)}
	\end{equation}
where $x_{0}$ and $x_{i}$ are measured with respect to the potential minimum within each potential well, and $x_{0}$ is a reference point that normally is chosen near the potential minimum ($x_{0}=0$).

Detailed numerical and analytic studies of feedback ratchets have been reported. The MIV scheme was found to perform less well than optimized, periodically flashed ratchets in the deterministic limit of large $N$ \cite{Cao:2004lr}, a drawback that can be remedied by introducing a force threshold  \cite{Dinis:2005fk,Feito:2006uq}. Realistic feedback implementations \cite{Craig:2008fr,Craig:2008ys} will need to take into account delays \cite{Feito:2007vn,Craig:2008ys,Feito:2008rr} in the implementation of feedback, as well as errors \cite{Feito:2007kx} in the measurement of particle positions and forces.   

	In this Letter, we implement and characterize an experimental feedback ratchet for small numbers of diffusive microbeads in water using optical line traps. We implement the MIV scheme and observe the predicted increase in $<\!\!v\!\!>$, as well as a small additional improvement using the MND scheme. Our experimental results compare well to results from Langevin dynamics simulations. We find that the MND scheme with feedback delay can be further improved by tuning the reference point $x_{0}$.

To create a line trap, a focused laser beam is scanned periodically and rapidly along a line such that a trapped sphere feels a time averaged optical potential.  If the scan speed and light intensity are kept constant, the sphere diffuses freely along the line, which has been used to study the entropic interactions  \cite{Crocker:1999fr} and single file diffusion of colloids \cite{Lutz:2004ul}. By modulating the light intensity appropriately within each scan, one can create a constant-force measuring tool for single molecule experiments \cite{Liesfeld:2003kx}, or a  time-dependent flashing ratchet potential \cite{FAUCHEUX:1995ys}.

We used an acousto-optic deflector (AOD) to scan a 1064 nm Nd:YAG laser back and forth at 2 kHz in one dimension transverse to the optical axis. To create an optical trap, we used an oil immersion microscope objective (100x, 1.4 NA, Leica Model No. 11566014)  to focus the beam into a microchamber. The $60~\um$ deep microchamber is composed of double sided tape, with a channel cut out, sandwiched between a coverslip and a microscope slide. Silica spheres, $0.9~\um$ diameter, in solution are injected into the microchamber with a syringe via a drilled input hole. The line trap system is used to trap up to  $N = 3$ spheres in a $17.4~\um$ long line containing ten ratchet periods ($L=1.74~\um$).

Brightfield images (200x50 pixels) are captured by a CCD camera at $130~\textrm{Hz}$ and are analyzed through LabVIEW  software to find the positions of the spheres, and a  feedback algorithm (MIV or MND) is then applied. The implementation of this  process takes about  $\tau = 5~\textrm{ms}$ per frame, much less than the time scale of about $300~\textrm{ms}$ for bead diffusion over a distance $aL$, which has been found to be the upper limit for tolerable feedback delay for $N = 1$  \cite{Craig:2008ys}.

\begin{figure}[!t]
\includegraphics{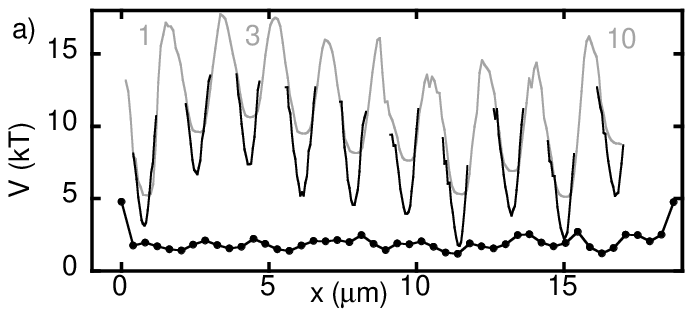}
\includegraphics{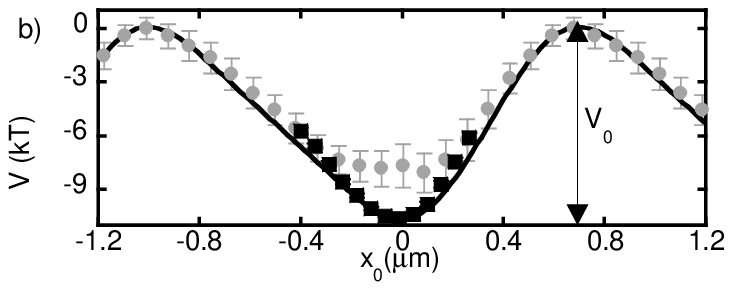}
\caption{\label{potential}  (a) The measured flat potential (black circles indicate experimental error) and the ratchet potential measured using Eq.~\eqref{population} and Eq.~\eqref{velocity} (black and gray lines, respectively, also in (b)).  (b) The measured ratchet potential, averaged over all ten periods, yields $a_{\textrm{eff}}=0.35 \pm 0.02$, and $V_{0}=10.6 \pm 0.3~kT$, compared to $V_{\textrm{eff}}(x)$ (black line) calculated from Eq.~\eqref{effective}, using $a=0.2$, $R = L/2$.}
\end{figure}

To define a potential shape, a periodic signal is sent  to the AOD via a function generator.  A potential minimum (maximum) is created by higher (lower)  intensity light.  However, because of angularly non-uniform AOD output and limits in AOD resolution, the resulting output light from the AOD will not correspond precisely to the input signal. To characterize the flat potential  (Fig.~\ref{potential}), we obtained statistics on the free diffusion of particles and determined the potential from
	\begin{equation}
	\label{population}
	V\left( x \right) =  - kT\ln \left( {\frac{{N\left( x \right)}}{{N_{total} }}} \right)
	\end{equation}
where $k$ is Boltzmann's constant, $T$ is temperature, $N(x)$ is the number of data points at position bin $x$, and $N_{total} \approx 10^5$ is the total number of data points. By iteratively measuring the effective potential and making adjustments to the input function we achieved a potential that was flat within one $kT$ and stable over the course of weeks.

Population sampling in equilibrium (Eq.~\eqref{population}) provides little data near the maxima of the ratchet potential when $V_{0}$ is more than a few $kT$. In those regions we use
\begin{equation}
	\label{velocity}
	V\left( x \right) =  - \gamma \int_0^x {\bar{v}\left( {x'} \right)} dx'
	\end{equation}
where $\gamma = kT/D$ is the drag coefficient of the microsphere ($D=0.44~\um^{2}/s$ as measured), and $\bar{v}\left( {x'} \right)$ is the ensemble average of the instantaneous velocity of the beads at position $x'$. To characterize the ratchet potential we use both Eq.~\eqref{population} and Eq.~\eqref{velocity} and merge the results by matching the slopes near the potential minimum, where both methods can be assumed to be reliable. We then determine the potential experienced by the beads as shown in Fig. \ref{potential}, where $V_{\textrm{lin}}$ with $a=0.2$ was used as an input to the AOD.

A finite-size sphere of radius $R$ spatially averages over a given one-dimensional ratchet potential $V(x)$ and experiences the effective potential  
\begin{equation}
\label{effective}
 V_{\textrm{eff}}(x,R)=\frac{3}{4\pi R^{3}}\int_{x-R}^{x+R}V(x')S(x') \, \mathrm{d}x'
\end{equation}
where $S(x')$ is the cross-sectional area perpendicular to the ratchet direction of the bead at position $x'$.  The effective potential calculated for our experimental parameters is shown in (Figs.~\ref{ratchetdiagram},\ref{potential}), and agrees well with the measured potential when $V(x')=V_{lin}$ with $a=0.2$ is used in Eq.~\eqref{effective} and when the result is scaled to match the measured $V_0$ (Fig.~\ref{potential}).

\begin{figure}[!ht]
\includegraphics{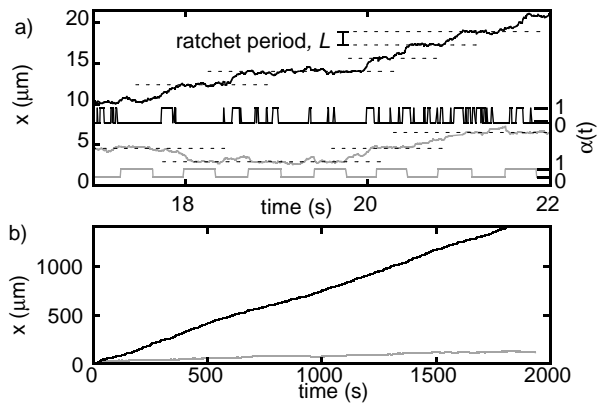}
\caption{\label{short} (a) Top black: particle trajectory for feedback-controlled ratcheting. Bottom black: $\alpha(t)$ for feedback trajectory. Gray: particle trajectory and $\alpha(t)$ for periodic ratcheting. (b) Longer time series of the same trajectories, generated by stitching together shorter runs.}
\end{figure}

To characterize the performance of our ratchet we must allow the possibility of back stepping. For this reason, measurements were taken by starting the microsphere at the minimum of period 3 (Fig.~\ref{potential}(a)), and the motion of the microsphere in response to thermal noise and ratchet flashing was recorded until the sphere reached the minimum of period 1 or 10.  Data collection was then halted and the microsphere was moved back to the starting position. This process was repeated until approximately $2\times10^5$ data points were collected, corresponding to a typical total displacement of more than $400~\textrm L$. Fig.~\ref{short} shows a sample trajectory obtained by periodically flashing the ratchet near its experimentally determined optimal period of $700~\textrm{ms}$, resulting in a time-averaged mean velocity of  $<\!\!v\!\!> = 0.068\pm0.01~\um/\textrm s$.  Using a mirrored potential we find  $<\!\!v\!\!> = -0.074\pm0.01~\um/\textrm s$, establishing  a small negative bias of less than $10~\textrm{nm/s}$. 

\begin{figure}[!t]
\includegraphics{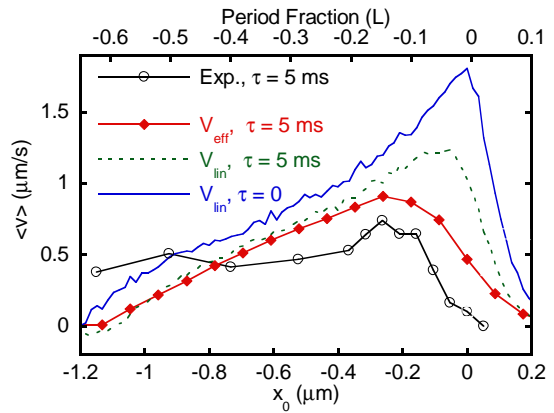}
\caption{\label{x0} (Color online) Average velocity $<\!\!v\!\!>$ as a function of $x_{0}$ for $N=1$ and MND. From the top: Simulated data using $V_{\textrm{lin}}$,  $a = 0.33$, $\tau = 0$ (solid blue curve); $V_{\textrm{lin}}$,  $a = 0.33$, $\tau = 4~\textrm{ms}$ (dashed green curve);  $V_{\textrm{eff}}$ and  $\tau = 5~\textrm{ms}$ (red diamonds); and experimental data $\tau = 5~\textrm{ms}$ (open black circles). }
\end{figure}

We first present  feedback control experiments using a potential of strength $V_{0}=10~kT$ for the case $N =1$, where the MIV scheme is identical to the MND scheme when $x_{0} = 0$. However, for the MND scheme, it is not initially clear whether  $x_{0} = 0$ is the optimal choice, and in Fig.~\ref{x0} we show the $<\!\!v\!\!>$ obtained using the MND feedback algorithm as a function of $x_{0}$. We find a maximum $<\!\!v\!\!>$  of $0.7~\um/\textrm s$, an order of magnitude larger than the maximum velocity found for periodic flashing, in agreement with predictions  for $N = 1$ \cite{Cao:2004lr, Craig:2008fr}.

An initially unexpected finding is that this peak velocity occurs for $x_{0} \approx - 0.3~\um$, $0.15~L$  to the left of the minimum.  To better understand the roles that experimental feedback delay and the precise effective potential shape may play in this behavior, we performed simulations as follows \cite{Craig:2008fr}. Particle trajectories were calculated by the overdamped Langevin equation
\begin{equation}
\begin{gathered}
\label{langevin}
\gamma \dot{x_i}=\alpha(t) F(x_{i}(t))+\xi_{i}(t)
\end{gathered}
\end{equation}
where $\xi_{i}(t)$ is a random number chosen from a Gaussian white noise distribution that has zero mean and correlation $\left<\xi_{i}(t)\xi_{j}(t')\right>=2\gamma kT \delta_{ij}\delta(t-t')$, and the external force is $F(x)=- \nabla V(x)$. Using for $V(x)$ the piecewise linear potential $V_{\textrm{lin}}(x)$ with $a = 0.33$ and no feedback delay, the simulations predict a maximum ratchet velocity for $x_{0} = 0$ (Fig. \ref{x0}), where $x_{0} = 0$ is always taken at the potential minimum. To include a feedback delay $\tau$ in the simulations, $\alpha(t)$ in Eq.~\eqref{langevin} is replaced by an effective delayed value, $\alpha(t-\tau)$. For $\tau = 5~\textrm{ms}$ (equal to the experimental value), the simulations produce a somewhat reduced peak velocity at  $x_{0} \approx - 0.1~\um$ for the piecewise linear potential (Fig. \ref{x0}). When using the same time delay and $V_{\textrm{eff}}(x)$ (Eq.~\eqref{effective}), we find that the peak velocity is shifted even further to negative $x_{0}$, to approximately the same position as in the experiment.

\begin{figure}[!t]
\includegraphics{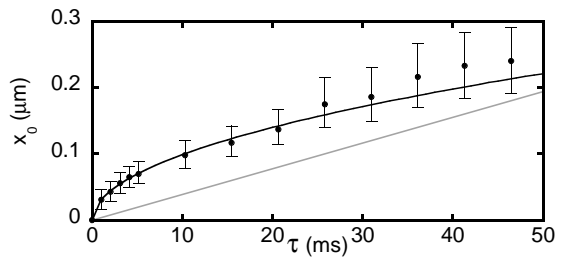}
\caption{\label{xnotfit}$x_0$ positions (circles) giving maximum $<\!\!v\!\!>$ from curves such as the inset of Fig.~\ref{x0improve} for various $\tau$.  For small $\tau$ these points fit well to a diffusive curve ($x=\sqrt{2D\tau}$, black line). Gray line: $x=-F \tau /\gamma$.}
\end{figure}

To understand why finite feedback delay can introduce a shift of the maximum $<\!\!v\!\!>$ to negative $x_{0}$ values, consider a situation where the potential is on and a particle is drifting down the long slope of the ratchet. To achieve optimal $<\!\!v\!\!>$, the potential should turn off at the instant when the particle reaches the potential minimum \cite{Cao:2004lr}. For finite $\tau$ it is therefore advantageous to anticipate the arrival of the particle at the potential minimum, and to trigger the potential to be turned off at a position that is passed by the particle on average at a time $\tau$ earlier. To test this explanation, we show in Fig.~\ref{xnotfit} the $x_{0}$ that produces maximal  $<\!\!v\!\!>$ as a function of $\tau$ from simulations using  $V_{\textrm{lin}}(x)$ with $a=0.33$. Also shown are the distance $x_{0}=-F\tau/\gamma$ a particle on the long slope would drift during $\tau$, and the average distance  $x_0=\sqrt{2D\tau}$ a particle would diffuse during $\tau$. For short $\tau$, where transport is mainly diffusive, the optimal choice of $x_{0}$ is indeed closely approximated by the diffusive prediction.

\begin{figure}[!t]
\includegraphics{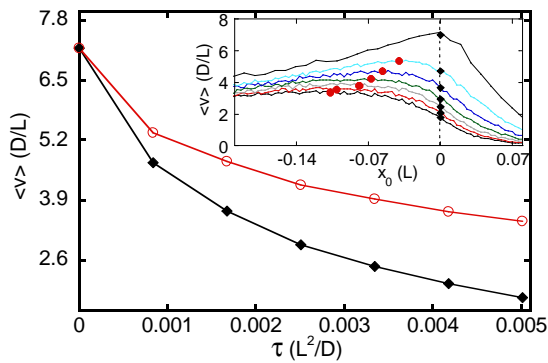}
\caption{\label{x0improve} (Color online) Simulated data using $V_{lin}$ and $a = 0.33$. Black diamonds are from the MIV and MND scheme for $x_0 = 0$. Open red circles are the maximum  $<\!\!v\!\!>$ obtained using MND and by tuning $x_0$ for each value of $\tau$. Inset: $<\!\!v\!\!>$ as a function of $x_0$ for $\tau =  0$ to $0.005~L^2/D$ (top to bottom)}
\end{figure}

An important implication of Fig. \ref{x0} is that existing feedback algorithms can be improved to better tolerate an implementation delay $\tau$, by introducing a spatial trigger  $x_{0} < 0$ that anticipates the arrival of particles at the potential minimum. Fig.~\ref{x0improve} illustrates this for the case $N = 1$. The lower curve shows the data reported in the seminal papers \cite{Feito:2007vn,Craig:2008ys} for the MIV scheme (equivalent to the MND scheme for  $x_{0} = 0$). The upper curve shows the optimal $<\!\!v\!\!>$ found by simulating the MND scheme and varying $x_{0}$ (see inset), demonstrating a substantial improvement for finite $\tau$. In simulations not shown here we find that for $N > 1$ the optimal MND scheme requires a finite $x_{0}$ even when $\tau = 0$. A future systematic evaluation of using spatial triggers in feedback algorithms for varying $N$ and $\tau$ is likely to lead to further improvements over existing algorithms.

\begin{figure}[!b]
\includegraphics{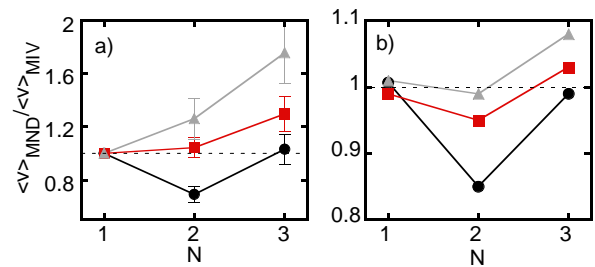}
\caption{\label{ratio} Ratio of $<\!\!v\!\!>$ obtained for MND and MIV methods for various $N$ and $V_{0}= 10kT, 30kT, 40kT$ from bottom to top. (a) Experimental data for optimal $x_0$. (b) Simulated data using  $V_{\textrm{lin}}$,  $a = 0.33$, $\tau = 0$ and optimal $x_0$ for MND. }
\end{figure}

For small $N > 1$ it has been predicted that the MND scheme can yield a moderately higher $<\!\!v\!\!>$ than the MIV scheme  \cite{Craig:2008fr}. To test this prediction, we performed experiments for $N=2,3$. To avoid any effects from particle-particle interactions, data points recorded when more than one bead was within any given ratchet period were excluded.  For $N > 3$ this happens so frequently that sufficient data cannot be collected. In Fig.~\ref{ratio}(a) we show experimental results for the ratio of average velocities obtained using the two schemes. In each case,  $x_{0}$ was varied, and the maximum $<\!\!v\!\!>$ was used in Fig ~\ref{ratio}(a). In agreement with simulations (Fig ~\ref{ratio}(b)) that used $V_{\textrm{lin}}(x)$ with $a = 0.33$, $\tau = 0$ and optimal $x_{0}$ for MND, the MND scheme performs better than the MIV scheme for $N = 3$ and $V_{0} \geq 30 kT$. A difference not shown here is that in the simulation $<\!\!v\!\!>$  increases with $V_{0}$ in all cases, whereas in the experiment we see an unexplained overall drop between $V_{0} = 30 kT$ and $40 kT$. This may be due distortions in $V(x)$ at high $V_0$.

We have demonstrated diffusive particle motion in customized, time dependent potentials with fast feedback. In the future this system can also be used to realize sorting devices akin to Maxwell's demon, or to study  particle-particle interactions in ratchets.

\begin{acknowledgments}
We acknowledge useful discussions with J. M. R. Parrondo and F. J. Cao. This research was supported by the National Science Foundation under CAREER Grant No. 0239764, NSF IGERT Grant No. DGE-0549503, and a grant from the Engineering and Technology Industry Council of Oregon.

\end{acknowledgments}




\end{document}